\documentclass[a4paper,fleqn,usenatbib]{mnras}

\usepackage{newtxtext,newtxmath}

\usepackage[T1]{fontenc}
\usepackage{ae,aecompl}
\usepackage{tablefootnote}   
\usepackage{booktabs,caption,fixltx2e}
\usepackage[flushleft]{threeparttable}

\usepackage{graphicx}	
\usepackage{amsmath}	
\usepackage{amssymb}	

\title[Do BL Lac Objects and FR I radiogalaxies  inhabit the same galaxy environment?]{
Do BL Lac Objects and  FR I radiogalaxies   inhabit the same galaxy environment?\\
}

\author[A. Sandrinelli, R. Falomo, A. Treves]{A. Sandrinelli$^{1},
$\thanks{E-mail: asandrinelli@yahoo.it }
R. Falomo$^{2}$, A. Treves$^{1,3}$\\
\footnotemark[1]\thanks{This work.....}\\
$^{1}$Universit\`a degli Studi del$\l'$Insubria, Via Valleggio 11, I-22100 Como, Italy\\
$^{2}$INAF -  Istituto Nazionale di Astrofisica, Osservatorio Astronomico di Padova, Vicolo dell\ÕOsservatorio 5, I-35122 Padova, Italy\\
$^{3}$INAF -  Istituto Nazionale di Astrofisica, Osservatorio Astronomico di Brera, Via Emilio Bianchi 46, I-23807 Merate, Italy\\
}

\date{Accepted XXX. Received YYY; in original form ZZZ}

\pubyear{2015}

\begin{document}
\label{firstpage}
\pagerange{\pageref{firstpage}--\pageref{lastpage}}
\maketitle

\begin{abstract}
We investigate the environments of galaxies around BL Lac objects and FR I radiogalaxies, 
the alleged parent populations of misaligned sources.
We compare the environment of a sample of 50 BL Lac objects at 0.1$<$z$<$0.33
with  that of a sample  of 90 FR I galaxies  at 0.1$<$z$<0.15$. 
The galaxy environment  is estimated using  the SDSS images in the i-band.
We find  that  the  galaxy excess density within 0.5 Mpc around of FR I  radiogalaxies is 
a factor $\sim$ 2 larger than that around BL Lacs.
This implies a reconsideration of the parent population of BL Lac objects.

  \end{abstract}




\section{Introduction}

   BL Lac objects (BLLs) are a class of sources originally characterized by 
   significant and rapid flux variability, high
polarization, and absence or weakness of spectral lines. In a seminal paper 
\cite{Blandford1978}  suggested that BLLs should be explained by relativistic jets 
pointed close to the observer direction. The parent population was soon after 
proposed as Fanaroff-Riley I radiogalaxies \citep[FRIs,][]{Fanaroff1974}, 
mainly on the basis of the extended radio emission, and the expected luminosity 
function assuming highly beamed sources \citep{Urry1994}. 
The relativistic beaming was directly confirmed by the detection of 
superluminal velocity in the inner radio structures of some BLLs 
and also by the more recent  observation of $\gamma$-ray emission \citep{Madejski2016}.
In fact the sources, with dimensions constrained by rapid variability, are transparent
to photon scattering of high energy gamma rays only if Lorentz contraction 
is accounted for. This picture, which consented to insert quite naturally 
BLLs  in the unified model of AGNs developed in the nineties
 \citep[see e.g.][]{Antonucci1993,Urry1995}, 
has essentially survived for about 30 years, even if enriched in various aspects, 
as for instance the discovery that the weakness of spectral lines should be attributed
not only to the relativistic enhancement of the continuum, but in part to be intrinsic.

The  contribution of  the Hubble Space Telescope exploration of the host galaxies has been 
substantial, showing  that they are giant ellipticals of magnitude M$_{R}$ $\sim$ -22.8, 
undistinguishable from unperturbed inactive ellipticals \citep[][]{Urry2000,Falomo2000,Sbarufatti2005}.

 The study of the clustering around BLLs  was originally based on 
 few objects known to belong to the class, and showed that BLLs 
are generally surrounded by modest galaxy groups, though some
 exceptions were noted
\setcitestyle{notesep={ },round,aysep={},yysep={;}}
\citep[see e.g. ][ for an overview]{Fried1993,Pesce1994,Wurtz1997,Falomo2014}.

The systematic survey in the GeV band in the last decade by the \textit{Fermi} 
satellite demonstrated that the extragalactic sky is dominated by BLLs 
and greatly increased the number of known ones, which includes now 
 more than a thousand objects
 \setcitestyle{notesep={; },round,aysep={},yysep={;}}
 \citep[][and references therein]{Padovani2017,Acero2015}.
  At the same time the Sloan Digital Sky Survey (SDSS) has covered a large fraction 
 of the sky in various filters, so that clustering of galaxies around many BLLs 
 can be studied in a renovated perspective with respect to some 20 years ago.

In this letter we select a sample of $\sim$ 50 BLLs at 0.1$<$z$<$0.33 belonging 
to the SDSS DR14 and search and study the surrounding environment. 
We then compare with the case of $\sim$ 200 inactive ellipticals from the SDSS  
matched in redshift and with the typical luminosity of the BL Lac host galaxies.  
Using the same procedure we consider also a sample of ~90 FRIs, 
again belonging to the SDSS, in the interval 0.1$<$z$<$0.15. 
The results are discussed, with particular emphasis 
on the comparison and contrast of galaxy clustering around BLLs and FRIs.

  \begin{figure}
\centering
\hspace{0.11cm}
\includegraphics[trim=-2cm 1cm 0.0cm 0.0cm,clip,width=1.5\columnwidth]{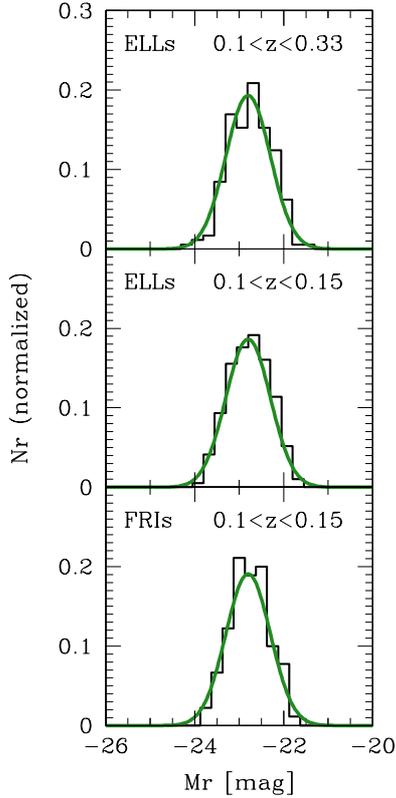}
\caption{ 
\label{3histM} 
\textit{From top to bottom}: Absolute magnitude distributions in r-band 
 for the  inactive elliptical galaxies   at 0.1$<$z$<$0.33 
and at  0.1$<$z$<$0.15, and for the  FR I radiogalaxies at 0.1$<$z$<$0.15 in our samples.
The gaussians represent the typical luminosity distribution of the BL Lac 
host galaxies derived from a sample of HST observations in \citet{Sbarufatti2005}.
}  
\end{figure}

\begin{figure}
\centering
\includegraphics[trim=0.3cm 1cm 0.0cm 0cm,clip,width=0.99\columnwidth]{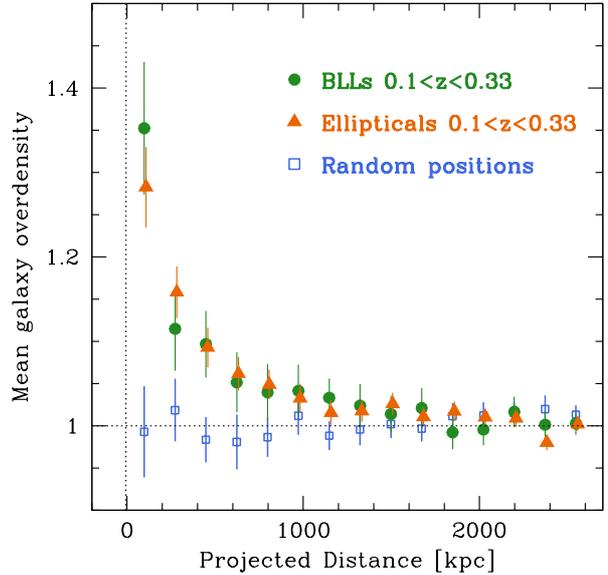}  
\caption{ 
\label{O} 
Average  overdensity of galaxies  with magnitude brighter than  mi$_{50\%}$ (see text)
as a function of the projected 
distance from  BL Lac objects (green filled circles), inactive  ellipticals
 (orange filled triangles) and  random positions  (blue open squares). 
Both  BL Lacs and inactive ellipticals exhibit the same 
 galaxy overdensity at projected distance  $\lesssim$ 1 Mpc. 
 The errors are the standard errors of the mean \citep[see also][]{Sandrinelli2018}.
}  
\end{figure}

\section{The samples}\label{thesamples}

We searched for BLLs suitable for investigating their environment using
the SDSS images and archive data. 
The selection of the dataset is based on the BZCAT catalog of known Blazars, 
Ed. 5.0\footnote{ www.asdc.asi.it/bzcat/} \citep{Massaro2015}. 
It contains 3,561 objects, of which 1,059 sources are spectroscopically classified
 as BLLs. We selected 676 BLLs  with available  SDSS images. 

We aim to detect galaxies  belonging to the environment of BLLs 
up to a magnitude threshold in the i-band corresponding to 
M$_i$ $\gtrsim$ M*+2.5  \citep[M*=$-$21.74][]{Montero2009}.
To reach this goal at the completeness of 50\% 
\citep[][]{Capak2007}\footnote{Durham University Cosmology Group,
references and data in \texttt{http://astro.dur.ac.uk/$\sim$nm/pubhtml/counts/counts.html}},
we need to limit the selection to objects  photometrically classified as galaxies
up to z $\lesssim$ 0.33  (mean threshold magnitude i=21.9 $\pm$ 0.1 mag).  
 This value  is taken as redshift upper limit for the sources in this study.
   We also set a lower value to the redshift at z $=$ 0.1 in order to limit the size of the fields. 
   This excludes  4 targets.
 We consider only  BLLs  with reliable spectroscopic classifications and redshifts. 
 Targets with corrupted images in the close surrounding  
 area\footnote{We note that $\sim$ 30\% of the objects
shows an incomplete coverage or corrupted frames in the area of  
our investigation, which extents up to 20 arcmin from each source.} are discarded. 
The final sample is composed of 47 BLLs with an average redshift z$_{ave}$ = 0.240.
 
For a direct comparison of the BLL environment with that around inactive elliptical
 galaxies, we select   $\sim$ 200 inactive ellipticals 
matched in redshift  with our BLL sample and in luminosity with 
the BLL host galaxies distribution  reported in \cite{Sbarufatti2005}.
 The targets,  
 with spectra in the  SDSS archives,
are  drawn 
on the basis of the contribution of   de Vaucouleurs  profile
 with respect to the exponential profile (\texttt{p.fracdev$_i>$0.8)}
  from good photometry (CLEAN=1)  galaxies  (\texttt{type = 3}). 
For further comparisons among the  environments (see Section \ref{theanalysis}),
 we also built a number of samples of  $\sim$ 200  inactive elliptical galaxies
  in different ranges of redshift.  
The luminosities of all the samples of inactive ellipticals 
are encompassed  between -24.0  $\lesssim$ M$_r$ $\lesssim $ -21.5 mag
 with mean absolute magnitude $\sim$ -22.8 $\pm$ 0.5  (standard error of the mean =   0.03),
 see Figure \ref{3histM} and Sect. \ref{theanalysis}. 
  
To build a suitable sample of FR I radiogalaxies we used the dataset of  
 \cite{Capetti2017}, which contains 219 sources  at z$<$0.15,
 obtained by combining observations from the NVSS, FIRST, and SDSS surveys. 
Consistently with the case of inactive ellipticals we built a sample  of objects 
at z$>$0.1 appearing in the SDSS and matching in absolute magnitude 
with the BLL host galaxies, see Figure \ref{3histM}. 
Again for both the inactive ellipticals  and FRI samples, targets with damaged 
images in the investigated areas are removed from the selections.
A number of  90 FRI galaxies are finally assembled,
with absolute magnitudes -22.76 $\pm$  0.45. 

\section{Analysis}\label{theanalysis}

 We drew information from  SDSS DR14 catalogues, and  obtain position
  and i-band photometry  of each target   and of  all primary objects 
classified as galaxies (\texttt{type = 3}).   
Absolute magnitudes M$_i$ of galaxies, evaluated at the redshift of the target
from \texttt{modelMag} magnitudes, are  corrected for galactic extinction 
based on the SDSS values and are also \textit{K}-corrected  assuming the 
elliptical template \citep{Mannucci2001}.

To study the galaxy  environment we follow the procedure
 described by \cite{Sandrinelli2014,Sandrinelli2018}.
We start  by computing  the surface number density  $n_r$ of galaxies  
brighter than a magnitude threshold (see Sect. \ref{theresults}), 
observed inside projected circular annuli  of central radius $r$. 
The radius is evaluated at the redshift of the target
and increases by steps of fixed width. The target source is excluded. 
 We compare $n_{r}$ with  the surface number galaxy density $n_{bg}$   
 in the outer region  (from    $\sim$ 2.5 Mpc to $\sim$ 4.5 Mpc),
taken as the background value, by adopting two  parameters:
the galaxy overdensity  defined as  $O_r=n_r/n_{bg}$, and 
  the excess galaxy surface density $E_r=(n_r-n_{bg})$.

\begin{figure*}
\centering
\includegraphics[trim=0.3cm 0.9cm 0.0cm 0.2cm,clip,width=1.2\columnwidth]{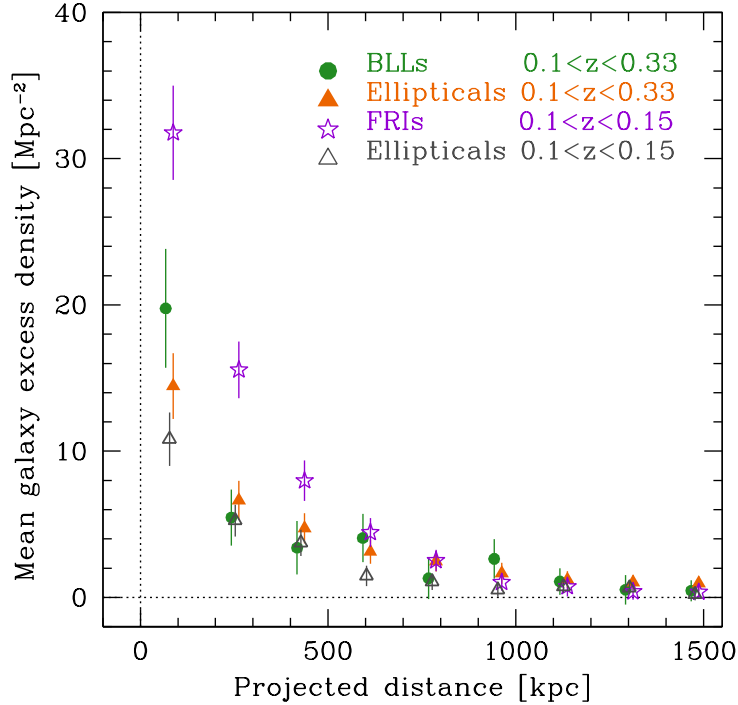}  
\caption{ 
\label{OellFRI} 
Average excess surface density of galaxies (detected in i-band, with magnitude 
threshold M*+2) as a function of the projected  distance from the BL Lac objects 
(green filled circles), inactive  ellipticals matched in redshift with the BLL sample
 (orange filled triangles), FRIs  (violet open stars) and  inactive ellipticals   
 (grey open triangles) matched in redshift with the FRI sample.
 }  
\end{figure*}

\begin{figure}
\centering
\includegraphics[trim=0.3cm 1cm 0.0cm 0.2cm,clip,width=0.99\columnwidth]{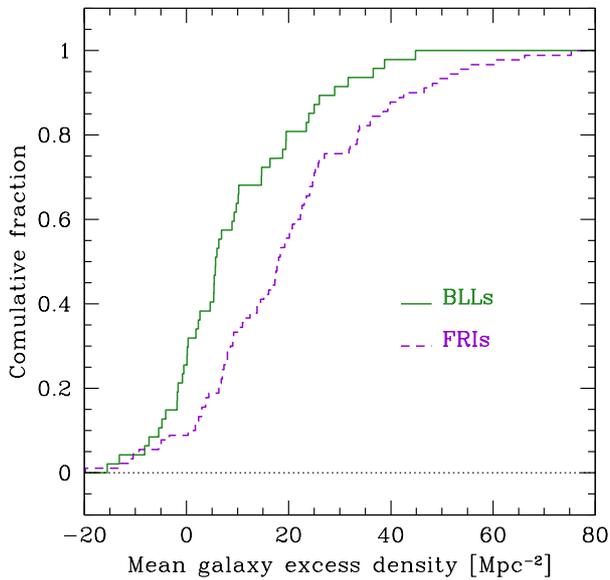}  
\caption{ 
\label{distr} 
The comparison of the cumulative distributions of excess galaxy density (within 350 kpc) 
of BL Lac objects (0.1$<$z$<$0.33) and FR I radiogalaxies (0.1$<$z$<$0.15). 
The two distribution are significantly different (see text).
}  
\end{figure}

\section{Results}\label{theresults}

The comparison of the distribution of galaxies around BLLs with the sample of
 inactive ellipticals that are matched both in redshift and galaxy luminosity shows
  that they inhabit the same galaxy environment (see Figure \ref{O}). 
The average  galaxy overdensity with respect to the background is detected up to 
a projected distance from the target of $\sim$ 1 Mpc and  within 0.5 Mpc  
is $O_r \sim$ 1.15. 
Furthermore, we perform a sanity test by searching for overdensity 
applying the identical analysis around random positions, 
obtained with shifts of 15 arcmin from  each BLL.
No evidence of overdensity is found  (see Figure \ref{O}) and therefore 
we consider  robust the result of similar clustering around BLLs and ellipticals.

We now aim to compare the galaxy clustering around BLLs and FRIs.
 Our samples are well matched in the luminosities,  but not in redshift, 
 and the BLL sample is too small to consider only the objects at z $<$0.15.
Therefore  we use as parameter  the excess densities $E_r$ and 
first compare  BLLs and ellipticals  matched in luminosity and redshift.
We use as threshold the apparent magnitude i, corresponding to $ M_i^*+2$
at the redshift  of each individual target. 
As illustrated in Figure \ref{OellFRI} we find on average a very good agreement, 
   consistently with the results on the overdensity.
  We have checked that this is the case also considering 5 different redshift
   intervals from z=0.1 to z=0.35, each containing $\sim 200$ ellipticals.  
    As expected, within the uncertainties we do not find any difference
     between $E_r$ evaluated  in the various redshift  intervals. 
     This is apparent in Figure \ref{OellFRI}, where the cases
   of ellipticals at 0.1$<z<$0.15 and at  0.1$<z<$0.33 are reported. 
 
The sample of FR I radiogalaxies is limited to 0.1$<$z$<$0.15 (see Sect. \ref{thesamples}),
 therefore we can only do a direct comparison with the sample of inactive ellipticals 
 in the same redshift range. 
We found that within $\sim$ 0.5 Mpc the excess density of the FR I radiogalaxies
 is a factor 2 larger than that of ellipticals (see Figure \ref{OellFRI}).  
Since we have shown that the galaxy environment of ellipticals does not change
 in the redshift range 0.1$<$z$<$0.35, we are confident that also the comparison 
 of environment between BLL and FR I radiogalaxies  is reliable. 
 The result is illustrated in Figure  \ref{OellFRI}.
  In particular,  within   0.5 Mpc the excess density is $\sim$ 13 Mpc$^{-2}$ 
  for the FRIs, while for the BLLs it is $\sim$ 6 Mpc$^{-2}$.

\section{Conclusions}\label{theconclusion}.

We find that the galaxy environment of low redshift BL Lac objects is indistinguishable 
from that of low redshift  inactive ellipticals of comparable galaxy luminosity. 
This result is similar to that found by  \cite{Karhunen2014} and \cite{Sandrinelli2018}
 who considered  large samples of low redshift  QSOs,
 finding an overdensity within 0.5 Mpc of $\sim$1.1.
 So it seems that in the local Universe BLLs, quasars and inactive ellipticals
 are similar in terms of surrounding galaxies.
 
We also find that the clustering of galaxies around radiogalaxies of FR I type
 is on average a factor $\sim$ 2  richer than that of BL Lac objects.
  In spite of the difference in the redshift distribution this difference 
  appears robust since no variation of galaxy environment is found 
  for inactive elliptical galaxies of same luminosity up to z $\sim$ 0.3 
  (see details in Sect. \ref{theresults}).
This result is further illustrated in Figure \ref{distr}, where we report the cumulative 
   distributions of the excess density  for the two classes of objects within the projected 
   distance 350 kpc.
   The medians are 5.7 and 13.6 Mpc$^{-2}$ for BLLs and FRIs, respectively, and 
     the KS test yields a p-value $<$ 4$\cdot$10$^{-4}$. 
     The population of BLLs is positioned at lower excess densities with respect FRIs.

      The different  clustering of galaxies around FRIs with respect to that of BLLs  
      was suggested some 20 years ago by \cite{Wurtz1997}.
     They considered $\sim$ 50 BL Lacs at z $<0.65$ and a similar number  
    of  FR I radiogalaxies at z $\lesssim$ 0.5,
       without performing any matching between the two classes
       in terms of redshifts and host galaxies luminosity distributions.
        
     The unambiguous distinction of FR I radiogalaxies and BLLs in terms 
     of clustering of galaxies in the immediate environments poses 
     a serious problem for the interpretation of the former class as the parent 
     population of the latter.
 A possible explanation of this is that BLLs have as parent population only 50\% 
 of the entire class of FR I radiogalaxies,  while the  remaining  objects with the higher 
 clustering are not connected with the BL Lacs.

\section*{Acknowledgements}
\ \ \ \ \ \ \ \ We thank Francesco Massaro for friendly and constructive conversations on the subjects
presented in this letter.
The use of the Sloan Digital Sky Survey facilities is gratefully  acknowledged. The SDSS web site is www.sdss.org.


\end {document}